Two-component description of dynamical systems that can be approximated by solitons: The case of the ion acoustic wave equations of Plasma Physics

Yair Zarmi
Jacob Blaustein Institutes for Desert Research
And Physics Department
Ben-Gurion University of the Negev
Midreshet Ben-Gurion, 84990
Israel

ABSTRACT

A new approach to the perturbative analysis of dynamical systems, which can be described approximately by soliton solutions of integrable nonlinear wave equations, is employed in the case of small-amplitude solutions of the ion acoustic wave equations of Plasma Physics. Instead of the traditional derivation of a perturbed KdV equation, the ion velocity is written as a sum of two components: elastic and inelastic. In the single-soliton case, the elastic component is the full solution. In the multiple-soliton case, it is complemented by the inelastic component. The original system is transformed into two evolution equations: An asymptotically integrable Normal Form for ordinary KdV solitons, and an equation for the inelastic component. The zero-order term of the elastic component is a single- or multiple-soliton-solution of the Normal Form. The inelastic component asymptotes into a linear combination of single-soliton solutions of the Normal Form, with amplitudes determined by soliton interactions, plus a second-order decaying dispersive wave. Satisfaction of a conservation law by the inelastic component and of mass conservation by the disturbance to the ion density is determined solely by the initial data and/or boundary conditions imposed on the inelastic component. The electrostatic potential is a first-order quantity. It is affected by the inelastic component only in second-order. The charge density displays a triple-layer structure. The analysis is carried out through third order.

PACS 02.30.IK, 02.30.Mv, 05.45.-a

MSC 35Q58, 41A58, 35Q51

Keywords Ion acoustic wave equations; Perturbed KdV equation; Normal Form; Soliton interactions.

## 1. Introduction

The KdV equation is an integrable approximation for several dynamical systems. It describes unidirectional solitons. The perturbed KdV equation (PKDVE) contains the terms that have been neglected in the approximation. Its derivation consists of two stages:

I. Expansion of the dynamical quantities of the original system in power series in a parameter.

II. Derivation of the PKDVE for the lowest-order approximation of one dynamical quantity.

The freedom in the expansion in either stage is substantial. In the traditional approach [1-7], where higher-order corrections are assumed to be differential polynomials in the zero-order approximation to the solution, the first-, second- and third-order corrections may contain up to 7, 23, and 74 monomials, respectively, with free coefficients.

The resulting PKDVE is asymptotically integrable if the expansion of the solution satisfies:

1) The zero-order approximation is the solution of an integrable Normal Form, which has the same soliton solutions as the unperturbed KdV equation, and merely updates soliton velocities.

2) Higher-order corrections are differential polynomials in the zero-order approximation.

The perturbation contains "obstacles to asymptotic integrability" if requirements 1 & 2 above cannot be satisfied simultaneously in some order. Obstacles emerge from second order onwards in the cases of the PKDVE [1-7] and the perturbed NLS equation [8, 9], and from first order - in the cases of the perturbed Burgers [10, 11], and mKdV [12] equations. Obstacles do not emerge in the single-wave case [6, 9, 13-16]. In the multiple-soliton case, they constitute a part of the inelastic soliton interactions. For a general solution, obstacles emerge unless the perturbation obeys specific constraints [5-9]. In the traditional analysis [1-7], obstacles spoil the simple structure of the zero-order approximation in the multiple-soliton case [5, 6, 8, 9]: The Normal Form ceases to be asymptotically integrable; soliton parameters develop time dependence; and the zero-order term loses the simple elastic scattering nature of solutions of the KdV equation.

Another hurdle is the emergence of contributions in the perturbative solution, which do not vanish asymptotically far from soliton trajectories in the multiple-soliton case. It is encountered in the traditional expansion [1-7] when obstacles to integrability exist [6, 9]. Non-vanishing terms of another type emerge in the analysis of the shallow-water equations: One is forced to incorporate in the solution a purely non-local third-order term, which does not vanish asymptotically far from soliton trajectories [7].

Small-amplitude solutions of the ion acoustic wave equations of Plasma Physics [17, 18] may not suffer from these problems. Admittedly, the existence of a second-order obstacle to integrability in the PKDVE was demonstrated in [6]. (The PKDVE was derived for the ion velocity, with specific choices made in the freedom in the expansion, through first order in [17, 19], and through second order in [6].) However, the numerical analysis of [19] showed that the solution of the full system is comprised of KdV solitons (with soliton parameters updated by the effect of the perturbation) plus a very small radiative tail that decays in time. This suggests that, perhaps, one may not be compelled to incorporate asymptotically non-vanishing terms in the perturbative solution for this system, and that obstacles to integrability need not spoil the simple physical picture.

A consistent expansion, which does not contain asymptotically non-vanishing non-local terms, can be obtained already in the traditional analysis [1-7] of the ion acoustic wave equations. This is shown in the first part of this paper. To overcome the effect of the unavoidable obstacles to integrability [6], a recently developed perturbation scheme [20], based on the observation that obstacles do not emerge in the single-soliton case [6, 9, 13-16], is then employed. The solution of the ion velocity is written as a sum of an elastic and an inelastic component. The original system is transformed into two evolution equations: An asymptotically integrable Normal Form for ordi-nary KdV solitons, and an equation for the inelastic component. The analysis is carried out through third order in the expansion parameter.

The elastic component is the full solution in the single-soliton case, and part of the solution in the multiple-soliton case. Its zero-order term is, respectively, a single- or a multiple-soliton solution of the Normal Form. The latter generates the same soliton solutions as the unperturbed KdV equation, and merely updates soliton velocities [1-7, 21-23]. Higher-order terms in the asymptotic series that describes the elastic component are differential polynomials in the zero-order approximation. They do not contain any non-local entities (see Appendix B). As a result, the elastic component does not induce updating of soliton parameters beyond their Normal-Form updating.

The inelastic component accounts for soliton interactions in the multiple-soliton case. The driving terms in its evolution equation are confined to the soliton-collision region. This results in a component that is localized along soliton trajectories. The lowest-order approximation to the inelastic component is a wave of solitons and anti-solitons. This lowest-order term tends asymptotically into a linear combination of zero-order single-soliton solutions of the Normal Form. Through third order, the freedom in the expansion allows for the construction of the inelastic contributions so that they all asymptote into linear combinations of single-soliton solutions. The only additional effect of the obstacles to integrability, which emerge from second order onwards, is a dispersive wave that was previously found in numerical experiments [24-26]. It decays exponentially away from soliton trajectories [20]. Hence, like the elastic component, the inelastic component does not induce changes in soliton parameters beyond their Normal Form updating.

The Normal Form analysis of the PKDVE is reviewed in Section 2. The ion acoustic wave equations are presented alongside with fundamental properties expected of localized wave solutions in Section 3. The traditional derivation for this system of the PKDVE through third order is presented in Section 4. Possibilities offered by the freedom in the expansion, as well as the un-avoidability of obstacles to asymptotic integrability are reviewed in Section 5. The new expansion scheme is presented in Section 6. Some consequences of physical interest are presented in Section 7. A summary is presented in Section 8. Technical details are given in Appendices A-C.

The two-component picture described above can be applied to a class of dynamical systems, the solutions of which can be approximated by localized soliton solutions of integrable non-linear wave equations. This includes, for instance, systems, which can be approximated by the KdV equation (e.g., the shallow-water problem) or the NLS equation (e.g., surface waves on a deep fluid layer, solitons in non-linear optics) and more.

**2. Normal Form analysis of the PKDVE – A review [1-7]**

The generic form of the PKDVE has the following structure through third order:

$$
\begin{aligned}
w_t &= 6 w w_1 + w_3 \\
&+ \varepsilon \left( 30\alpha_1 w^2 w_1 + 10\alpha_2 w w_3 + 20\alpha_3 w_1 w_2 + \alpha_4 w_5 \right) \\
&+ \varepsilon^2 \begin{pmatrix} 140\beta_1 w^3 w_1 + 70\beta_2 w^2 w_3 + 280\beta_3 w w_1 w_2 + 14\beta_4 w w_5 + 70\beta_5 {w_x}^3 \\ + 42\beta_6 w_1 w_4 + 70\beta_7 w_2 w_3 + \beta_8 w_7 \end{pmatrix} \\
&+ \varepsilon^3 \begin{pmatrix} 630\gamma_1 w^4 w_1 + 1260\gamma_2 w (w_1)^3 + 2520\gamma_3 w^2 w_1 w_2 + 1302\gamma_4 w_1 (w_2)^2 \\ + 420\gamma_5 w^3 w_3 + 966\gamma_6 (w_1)^2 w_3 + 1260\gamma_7 w w_2 w_3 + 756\gamma_8 w w_1 w_4 \\ + 252\gamma_9 w_3 w_4 + 126\gamma_{10} w^2 w_5 + 168\gamma_{11} w_2 w_5 + 72\gamma_{12} w_1 w_6 + 18\gamma_{13} w w_7 + \gamma_{14} w_9 \end{pmatrix} \\
&+ O\left(\varepsilon^4\right) \qquad\qquad \left( |\varepsilon| \ll 1 \quad , \equiv w_p \equiv \partial_x{}^p w \right)
\end{aligned}
\quad .(2.1)
$$

As in the case of perturbed ODE's [27-35], the Normal Form analysis of Eq. (2.1) entails two asymptotic expansions [1-7]. The first one is the power expansion of the approximate solution:

$$
w(t,x) = u(t,x) + \varepsilon u^{(1)}[u] + \varepsilon^2 u^{(2)}[u] + \varepsilon^3 u^{(3)}[u] + O\left(\varepsilon^4\right) \quad . \tag{2.2}
$$

The second expansion is the Normal Form, expected to govern evolution of the zero-order term, *u*:

$$
u_t = S_2[u] + \varepsilon \alpha_4 S_3[u] + \varepsilon^2 \beta_8 S_4[u] + \varepsilon^3 \gamma_{14} S_5[u] + O\left(\varepsilon^4\right) \quad . \tag{2.3}
$$

(See Appendix A for the definition of the symmetries, $S_n$.)

Eq. (2.3) is asymptotically integrable, and has the same soliton solutions as the unperturbed KdV equation [1-7, 20-22]. Denoting the wave number of a soliton by $k$, the Normal Form merely updates soliton velocities according to:

$$v = v_0 + \varepsilon \alpha_4 v_0^{\,2} + \varepsilon^2 \beta_8 v_0^{\,3} + \varepsilon^3 \gamma_{14} v_0^{\,4} + O\left(\varepsilon^4\right) \quad , \quad \left(v_0 = 4k^2\right) \ , \tag{2.4}$$

In Eq. (2.2), the higher-order corrections, $u^{(n)}$, are assumed to be differential polynomials $u$. Their allowed forms are given in Eqs. (B.2)-(B.4). The calculation is reduced to the determination of the unspecified coefficients in the polynomials.

Focusing on soliton solutions of Eq. (2.3), it is desirable to ensure localization around soliton trajectories also in the higher-order corrections in Eq. (2.2). In the multiple-soliton case, all the purely non-local terms that are allowed in Eqs. (B.2)-(B.4) do not vanish asymptotically far from soliton trajectories. To preserve the localized nature, one eliminates all these terms by setting

$$a_i = 0 \quad \left(5 \le i \le 7\right) \quad , \quad b_j = 0 \quad \left(14 \le j \le 23\right) \quad , \quad c_k = 0 \quad \left(42 \le k \le 74\right) \ . \tag{2.5}$$

Eq. (2.5) is assumed throughout the paper. As will be shown in the following, a consistent expansion is obtained. A detailed study shows that inclusion of the eliminated terms does not lead to the salvation of asymptotic integrability. (In the shallow-water problem, one cannot apply Eq. (2.5) to the third-order coefficients, $c_k$. A purely non-local term must be included; at least $c_{52} \neq 0$ [7].)

The first-order analysis yields the following values for the coefficients in $u^{(1)}$ [1-7]:

$$a_1 = \frac{5}{6}\left(-3\alpha_1 + 2\alpha_3 + \alpha_4\right) \ , \ a_2 = -\frac{10}{3}\left(\alpha_2 - \alpha_4\right) \ , \ a_3 = 0 \ , \ a_4 = \frac{5}{3}\left(-3\alpha_1 + \alpha_2 + 2\alpha_4\right) \ . \tag{2.6}$$

In the multiple-soliton case, one may run into an algebraic impasse in orders, $n \geq 2$: The coefficients of monomials in $u^{(n)}$ may not be determinable. It is then impossible to account for part of the perturbation in the dynamical equation obeyed by $u^{(n)}$. This part is an "obstacle to asymptotic integrability" [2-11]. Obstacles do not emerge only if specific combinations of the coefficients in the perturbation in Eq. (2.1) vanish. An $O(\varepsilon^2)$-obstacle does not emerge if [2-7]

$$\begin{aligned}\mu_2 = 5\left(3\alpha_1\alpha_2 + 4\alpha_2{}^2 - 18\alpha_1\alpha_3 + 60\alpha_2\alpha_3 - 24\alpha_3{}^2 + 18\alpha_1\alpha_4 - 67\alpha_2\alpha_4 + 24\alpha_4{}^2\right) \\ + 21\left(3\beta_1 - 4\beta_2 - 18\beta_3 + 17\beta_4 + 12\beta_5 - 18\beta_6 + 12\beta_7 - 4\beta_8\right) = 0\end{aligned} \quad . \tag{2.7}$$

If Eq. (2.7) is not obeyed, then, through second order, Eq. (2.3) is modified into [5, 6]

$$u_t = S_2[u] + \varepsilon\alpha_4 S_3[u] + \varepsilon^2\left(\beta_8 S_4[u] + \mu_2 R_2[u]\right) + O\left(\varepsilon^3\right) \quad . \tag{2.8}$$

Eq. (2.8) is not integrable owing to the presence $R_2[u]$, the unaccounted for part of the second-order perturbation. In the multiple-soliton case, the solution of Eq. (2.8) does not have the simple structure of the solution of the KdV equation [5, 6, 8, 9]. If Eq. (2.7) is obeyed, then obstacles emerge in the third-order calculation, unless three combinations of the $\alpha$-, $\beta$- and $\gamma$- coefficients of Eq. (2.1), given in Eqs. (C.1)-(C.3), vanish. When $u$ is a single-soliton solution, obstacles do not emerge even if Eq. (2.7) or its $O(\varepsilon^3)$ counterparts are not obeyed [5, 6, 16].

### 3. The ion acoustic wave equations

The ion acoustic wave equations represent a collision-less plasma made of cold ions and hot electrons in (1 + 1) dimensions. In terms of dimensionless quantities, they are [17-19]:

$$n_\tau + (n v)_\xi = 0 \quad , \tag{3.1}$$

$$v_\tau + \left(\frac{1}{2}v^2 + \varphi\right)_\xi = 0 \quad , \tag{3.2}$$

$$\varphi_{\xi\xi} = e^\varphi - n \quad . \tag{3.3}$$

Eqs. (3.1) and (3.2) are the continuity and momentum-conservation equations of Fluid Dynamics. Eq. (3.3) is the Poisson equation for the electrostatic potential, $\varphi(\tau, \xi)$. The electron density, ion density and ion velocity are $e^{\varphi}$, $n(\tau, \xi)$ and $v(\tau, \xi)$, respectively. $\tau$ is time, and $\xi$ is the horizontal spatial coordinate.

If the deviations of $n(\tau, \xi)$ and $v(\tau, \xi)$ from their asymptotic values ($n_0$ and $v_0$, respectively) vanish sufficiently rapidly for $|\xi| \to \infty$ at fixed time, then Eq. (3.1) implies mass conservation:

$$\frac{d}{d\tau}\int_{-\infty}^{+\infty}\left(n(\tau,\xi) - n_0\right)d\xi = 0 \quad . \tag{3.4}$$

In addition, Eq. (3.2) yields the overall change in the electrostatic potential, $\varphi(\tau, \xi)$:

$$\varphi(\tau,\xi \to +\infty) = -\frac{d}{d\tau}\int_{-\infty}^{+\infty}\left(v(\tau,\xi) - v_0\right)d\xi \quad . \tag{3.5}$$

**4. Traditional derivation [1-7] of PKDVE from ion acoustic wave equations [17-19]**

Considering disturbances with amplitude of $O(\varepsilon)$, $|\varepsilon| \ll 1$, one first rescales according to:

$$n(\tau,\xi) = N(\sigma,x), \quad v(\tau,\xi) = V(\sigma,x), \quad \varphi(\tau,\xi) = \Phi(\sigma,x) \quad \left(\sigma = \varepsilon^{3/2}\tau, \quad x = \varepsilon^{1/2}\xi\right) \quad . \tag{4.1}$$

Eqs. (3.1) – (3.3) become:

$$\varepsilon N_\sigma + (NV)_x = 0 \quad , \tag{4.2}$$

$$\varepsilon V_\sigma + V V_x + \Phi_x = 0 \quad , \tag{4.3}$$

$$\varepsilon \Phi_{xx} = e^{\Phi} - N \quad . \tag{4.4}$$

The stationary solutions of Eqs. (4.2) – (4.4) are

$$N_0 = 1 \quad , \quad V = V_0 \quad , \quad \Phi_0 = 0 \quad . \tag{4.5}$$

Assuming for $\Phi$ a zero-boundary condition at $x \to -\infty$, Eqs. (4.3) and (4.4) yield:

$$\Phi = -\,\varepsilon\partial_\sigma\int_{-\infty}^{x}\left(V - V_0\right)dx - \tfrac{1}{2}\left(V^2 - V_0^{\,2}\right) \quad , \quad N = e^{\Phi} - \varepsilon\Phi_{xx} \quad . \tag{4.6}$$

A consistent expansion can be obtained for $V_0 = \pm 1$. As the results of the analysis for the two values are identical, $V_0 = +1$ is adopted for the unidirectional solution. To derive the PKDVE through third order, one has to expand $V$ through $O(\varepsilon^4)$:

$$V = 1 + \sum_{n\geq 1} \varepsilon^n V^{(n)} \quad . \tag{4.7}$$

Substituting Eqs. (4.5)-(4.7) in Eq. (4.2), the latter yields:

$$\partial_\sigma V^{(1)} = -V^{(1)}\partial_x V^{(1)} + \tfrac{1}{2}\partial_x{}^3 V^{(1)} + \sum_{n\geq 1}\varepsilon^n\left(-\partial_x\left(V^{(1)}V^{(n+1)}\right) + \tfrac{1}{2}\partial_x{}^3 V^{(n+1)} - \partial_\sigma V^{(n+1)} + Q^{(n)}\right). \tag{4.8}$$

In Eq. (4.8), $Q^{(n)}$ are known functionals of $V^{(k)}$, with $1 \leq k \leq n$. For example,

$$Q^{(1)} = -\tfrac{1}{2}\partial_x{}^{-1}\left(\partial_\sigma{}^2 V^{(1)}\right) + \tfrac{1}{2}\left(V^{(1)}\right)^2 \partial_x V^{(1)} + 2\partial_x V^{(1)}\partial_x{}^2 V^{(1)} + V^{(1)}\partial_x{}^3 V^{(1)} + \partial_\sigma\partial_x{}^2 V^{(1)} \quad . \tag{4.9}$$

To transform Eq. (4.8) into the canonical form of Eq. (2.1), one rescales according to:

$$V^{(1)}(\sigma,x) = -3w(t,x) \quad , \quad V^{(n)}(\sigma,x) = -3w^{(n-1)}(t,x),\, n\geq 2 \qquad (t = 2\sigma) \quad . \tag{4.10}$$

Eq. (3.13) now becomes:

$$\begin{aligned} w_t &= 6ww_1 + w_3 \\ &\quad + \varepsilon\, J^{(1)}\left[\partial_t, w, w^{(1)}\right] + \varepsilon^2 J^{(2)}\left[\partial_t, w, w^{(1)}, w^{(2)}\right] + \varepsilon^3 J^{(3)}\left[\partial_t, w, w^{(1)}, w^{(2)}, w^{(3)}\right] + O\left(\varepsilon^4\right) \end{aligned} \tag{4.11}$$

$J^{(n)}$ are known differential polynomials in $w$ and $w^{(k)}$, $1 \leq k \leq n$. Once the polynomial expressions of Appendix B for $w^{(n)}$ are substituted in Eq. (4.11), and Eq. (4.11) is employed repeatedly so as to eliminate derivatives of $w$ with respect to time in its r.h.s., the perturbed equation becomes:

$$w_t = 6ww_1 + w_3 + \varepsilon\, K^{(1)}[w] + \varepsilon^2 K^{(2)}[w] + \varepsilon^3 K^{(3)}[w] + O\left(\varepsilon^4\right) \quad . \tag{4.12}$$

In Eq. (4.12), $K^{(n)}$ are differential polynomials in $w$. For example, the $O(\varepsilon)$ term is:

$$K^{(1)}[w] = 6a_3\left\{q^2 w w_1 + q\left(w^3 + w_1^2 + w w_2\right)\right\}$$
$$+ \left(3a_2 + 12a_3 + 6a_4\right)w^2 w_1 + \left(3a_2 - 3\right)w w_3 + \left(\frac{3}{2} - 12a_1 + 3a_2 + 6a_4\right)w_1 w_2 + \tfrac{3}{4}w_5 \quad .(4.13)$$

Like $K^{(1)}$, $K^{(n)}$ , for all $n \geq 1$, contain non-local terms, such as $q(t,x)$. For Eq. (4.12) to assume the form of Eq. (2.1), they must be eliminated. Through third order, this is achieved by the following coefficient assignments in Eqs. (B.2)-(B.4):

$$a_3 = 0 \quad , \tag{4.14}$$

$$b_3 = \frac{1}{2}a_2^2 \quad , \quad b_9 = -4b_6 \quad , \quad b_{12} = -6a_1 a_2 + a_2^2 + 2a_2 a_4 + 6b_2 \quad , \quad b_4 = b_7 = b_8 = b_{13} = 0 \quad , \tag{4.15}$$

$$c_3 = a_2\left(-\frac{1}{2}a_1 a_2 + b_2\right) \quad , \quad c_4 = \frac{1}{6}a_2^3 \quad , \quad c_7 = a_2 b_5$$
$$c_{14} = a_2\left(\frac{5}{4}a_2 - \frac{1}{2}a_1 a_2 - \frac{1}{2}a_2^2 - \frac{5}{4}a_4 + 2a_1 a_4 + a_2 a_4 - a_4^2 - \frac{1}{2}b_2 - b_5 - b_9 + b_{10} + \tfrac{1}{2}b_{13}\right)$$
$$+ \left(\frac{5}{4} - 2a_1 + a_4\right)b_5 - \frac{1}{2}c_{12}$$
$$c_{28} = a_2\left(\frac{21}{2}a_1 + 4a_1^2 - 3a_1 a_2 - 2a_1 a_4 - 10b_1 + 4b_2 + b_9\right) + \left(-\frac{21}{2} - 4a_1 + 2a_4\right)b_2$$
$$+ 10c_2$$
$$c_{29} = a_2\left(\frac{27}{2}a_1 + 14a_1^2 - 4a_1 a_2 - 6a_1 a_4 - 20b_1 + 4b_2 + b_9 + 2b_{10}\right) + \left(-\frac{27}{2} - 14a_1 + 6a_4\right)b_2$$
$$+ 20c_2 \quad .(4.16)$$
$$c_{30} = a_2\left(-6a_1^2 + a_2^2 + a_2 a_4 + 6b_2\right) \quad , \quad c_{31} = a_2\left(-6a_1^2 + \tfrac{1}{2}a_2^2 + a_2 a_4 + 6b_2\right)$$
$$c_{33} = \left(-6a_1 + a_2 + 2a_4\right)b_5 + 6c_5$$
$$c_{39} = a_2\left(\frac{45}{2}a_1 + 30a_1^2 - 12a_1 a_2 + a_2^2 - 18a_1 a_4 + 2a_2 a_4 - 30b + 12b_2 + b_5 + 3b_{13}\right)$$
$$+ \left(-\frac{45}{2} - 30a_1 + 18a_4\right)b_2 + 20c_2$$
$$c_6 = c_8 = c_9 = c_{10} = c_{11} = c_{13} = c_{15} = c_{16} = c_{17} = c_{18} = c_{19} = c_{20} = c_{21} = c_{22} = c_{23} = c_{24} = c_{32}$$
$$= c_{34} = c_{35} = c_{36} = c_{40} = 0$$

All other coefficients remain free. The coefficients in Eq. (2.1) now obtain the following values:

$$\alpha_1 = \frac{1}{10}a_2 + \frac{1}{5}a_4 \quad , \quad \alpha_2 = \frac{3}{10}\left(a_2 - 1\right) \quad , \quad \alpha_3 = \frac{3}{40} - \frac{3}{5}a_1 + \frac{3}{20}a_2 + \frac{3}{10}a_4 \quad , \quad \alpha_4 = \frac{3}{4} \quad , \tag{4.17}$$

$$
\begin{aligned}
\beta_1 &= \frac{1}{70}\left\{-9a_1a_2 + a_2^{\,2} + 2a_2a_4 + 9b_2 + b_5 + 3b_{13}\right\} \\
\beta_2 &= \frac{3}{140}\left\{3 - 6a_2 - 2a_4 - 14a_1a_2 + 4a_2^{\,2} + 14b_2 + 2b_5\right\} \\
\beta_3 &= \frac{3}{560}\left\{-3 - 3a_2 - 10a_4 - 56a_1a_2 + 8a_2^{\,2} + 12a_2a_4 - 8a_4^{\,2} + 48b_2 + 4b_5 - 8b_9 + 12b_{13}\right\} \\
\beta_4 &= \frac{3}{56}\left\{-6 + 5a_2 - 4a_1a_2 + 4b_2\right\} \\
\beta_5 &= \frac{3}{280}\left\{-9 - a_2 + 4a_4 - 24a_1a_2 + 4a_2^{\,2} - 16a_1a_4 + 4a_2a_4 + 24b_2 + 4b_5 - 8b_{10} + 8b_{13}\right\} \\
\beta_6 &= \frac{1}{112}\left\{-11 + 16a_1 + 20a_2 + 20a_4 + 32a_1^{\,2} - 24a_1a_{24} - 16a_1a_4 - 64b_1 + 8b_2 + 8b_9\right\} \\
\beta_7 &= \frac{3}{560}\left\{11 - 8a_1 + 30a_2 + 40a_4 + 112a_1^{\,2} - 32a_1a_2 - 48a_1a_4 - 160b_1 + 8b_9 + 16b_{10}\right\} \\
\beta_8 &= \frac{5}{8}
\end{aligned}
\quad . \quad (418)
$$

Using Eqs. (B.1) and (B.4) for $w^{(3)}$, the analysis yields expressions for $\gamma_j$, the coefficients in the $O(\varepsilon^3)$-perturbation in Eq. (2.1). The expressions for $\gamma_1$ and $\gamma_{14}$ are shown as examples:

$$
\begin{aligned}
\gamma_1 &= \frac{3}{28}a_1a_2 + \frac{1}{7}a_1^{\,2}a_2 - \frac{17}{420}a_1a_2^{\,2} + \frac{1}{420}a_2^{\,3} - \frac{1}{15}a_1a_2a_4 + \frac{1}{210}a_2^{\,2}a_4 \\
&\quad - \frac{1}{7}a_2b_1 + b_2\left(-\frac{3}{28} - \tfrac{1}{7}a_1 + \frac{17}{420}a_2 + \frac{1}{15}a_4\right) + b_5\left(-\frac{2}{105}a_1 + \frac{1}{210}a_2 + \frac{1}{210}a_4\right) \\
&\quad + \frac{1}{7}c_2 + \frac{2}{105}c_5 + \frac{1}{420}c_{14} \\
\gamma_{14} &= \frac{35}{64}
\end{aligned}
\quad . \quad (4.19)
$$

**5. Successes & difficulties in exploitation of freedom in traditional expansion**

The first success of the traditional expansion is that the freedom in the expansion allows for the construction of a solution that obeys Eq. (2.5). It does not contain purely non-local terms that do not vanish away from soliton trajectories. Hence, with the assignment, Eqs. (4.14)-(4.19), for the coefficients in Eqs. (B.2)-(B.4), the solution has the potential to obey Eqs. (3.4) and (3.5) through $O(\varepsilon^3)$, provided that $w(t,x)$ of Eq. (2.2) vanishes away from soliton trajectories (for $|x| \to \infty$ at fixed $t$). This requires that the zero-order approximation, $u(t,x)$, has the same asymptotic behavior.

Next, some simplification of the resulting solution is possible. The $O(\varepsilon)$-perturbation in Eq. (2.1) is conservative for any values of $\alpha_i$, $1 \le i \le 4$. Hence, if $w$ vanishes sufficiently rapidly for $|x| \to \infty$ at fixed time, $t$, then the resulting Eq. (2.1) conserves mass through $O(\varepsilon)$:

$$\frac{d}{dt}\int_{-\infty}^{+\infty} w(t,x)dx = O(\varepsilon^2) \quad . \tag{5.1}$$

To extend the validity of Eq. (5.1) through second order, the following relation must hold:

$$\beta_3 = (\beta_2 + \beta_5)/2 \quad . \tag{5.2}$$

One can satisfy Eq. (5.2), because there are many free $b$-coefficients in Eqs. (4.18) . Similar freedom exists in the choice of $c$-coefficients, examples of which are given in Eq. (4.19). Hence, Eq. (5.1) can be extended into

$$\frac{d}{dt}\int_{-\infty}^{+\infty} w(t,x)dx = O(\varepsilon^4) \quad . \tag{5.3}$$

As in the case of Eqs. (3.4) and (3.5), the validity of Eq. (5.3) depends on the asymptotic behavior of the zero-order approximation, $u$.

Finally, through $O(\varepsilon)$, the PKDVE can be made into an asymptotically integrable Normal Form (see Eq. (2.3)): It is possible to convert the $O(\varepsilon)$ perturbation in Eq. (2.1) into the symmetry $S_3$ (see Appendix A) by choosing the coefficients $a_i$ so that in Eqs. (4.17) one has [1-7]

$$\alpha_1 = \alpha_2 = \alpha_3 = \alpha_4 \quad . \tag{5.4}$$

Then, through first order, the solution for $w$ will have the same form as the solution of the KdV equation. Hence, it will vanish away from soliton trajectories.

However, the emergence of an $O(\varepsilon^2)$-obstacle to asymptotic integrability is unavoidable [6]. Eq. (2.7), the condition for the absence of an obstacle, is not satisfied. Using Eqs. (4.17) and (4.18) in Eq. (2.7), the contributions of the free *a*- and *b*- coefficients cancel out, and

$$\mu_2 = -(33/20) \quad . \tag{5.5}$$

Obstacles are unavoidable also in $O(\varepsilon^3)$. Substituting the expressions for the $\alpha$-, $\beta$- and $\gamma$- coefficients in the expressions for $\mu_{3i}$, $1 \le i \le 3$ (see Appendix C), all $\mu_{3i}$ do not vanish:

$$\mu_{31} = (451/240) \quad , \quad \mu_{32} = -(5711/960) \quad , \quad \mu_{33} = -(5711/4800). \tag{5.6}$$

Owing to the existence of obstacles, it is impossible to make Eq. (2.1) into the integrable Normal Form, Eq. (2.3) beyond $O(\varepsilon)$. As a result, all the simplifications discussed above cannot be implemented beyond $O(\varepsilon)$ in the traditional analysis. In particular, Eq. (5.3), and, more important, Eqs. (3.4) and (3.5) cannot be obeyed. The reason is that the zero-order term, $u$, now solved for from Eq. (2.8), does not vanish asymptotically far from soliton trajectories [5, 6, 8, 9].

**6. Two-component description in multiple soliton case and the role of special polynomials**

In the single-soliton case, it is possible to obtain a consistent expansion that does not include *any* non-local terms in the differential polynomials for $w^{(n)}$ (Eqs. (B.2)-(B.4)). Moreover, obstacles to asymptotic integrability do not emerge in this case [5, 6, 8, 9]. Both statements can be checked directly by analyzing Eqs. (3.1)-(3.3) in the single-soliton case. In this paper, they will become self evident from the manner in which the solution is constructed.

In the multiple-soliton case, the solution contains inelastic terms from first order onwards. Some have closed-form expressions that contain non-local entities [36]. Others are generated by obstacles to integrability, and cannot be written as differential polynomials. The obstacles spoil the simple structure of the zero-order approximation to the solution [5, 6, 8, 9]; it may not vanish away from soliton trajectories. To avoid spoiling the simplicity of the zero-order approximation, and to account for inelastic soliton interactions, the expansion scheme, originally applied to the PKDVE in [20], is applied directly to the physical system under consideration. This scheme evades the need to derive a PKDVE that suffers from obstacles to asymptotic integrability.

The ion velocity is written as a sum of an elastic and an inelastic component. The analysis of Section 4 is unaltered up to Eq. (4.10), which is replace by

$$V^{(1)}(\sigma,x) = -3w(t,x)\ ,\ V^{(2)}(\sigma,x) = -3\left(w^{(1)}[w] + \eta(t,x)\right)\ ,\ V^{(n)}(\sigma,x) = -3w^{(n-1)}[w]\,,\, n \geq 3 \quad (t = 2\sigma) \quad . \tag{6.1}$$

The freedom in the expansion will be exploited so as to achieve the following:

1. The lowest-order term, $w(t,x)$, is to be the solution of the Normal Form, Eq. (2.3). Hence, it describes a KdV-type multiple soliton solution, with soliton velocities updated by the effect of the resonant terms included in the Normal Form through Eq. (2.4).
2. $w^{(n)}[w]$, $n \geq 1$, contribute only to the elastic component. They are to have the same structure in the single- and multiple-soliton cases: purely local polynomials in $w$ (i.e., constructed out of powers of $w$ or its spatial derivatives).
3. $\eta(t,x)$ is to account for all inelastic soliton interactions in all orders: those that can be expressed as differential polynomials in $w$, as well as the effect of obstacles to asymptotic integrability.

The first-order analysis is presented in detail. In view of the cumbersome nature of the calculations, second-order results are presented with less detail, and only final results – in the third-order.

### 6.1 First-order analysis

$w^{(1)}$, the first-order contribution to the elastic component, must not contain *any* non-local terms, as they generate inelastic contributions [36, 12, 20]. Hence, of the seven monomials allowed in Eq. (B.2), only the two purely local ones are included:

$$w^{(1)} = a_1 w_2 + a_4 w^2 \quad . \tag{6.2}$$

With Eq. (6.2) for $w^{(1)}$, and the inclusion of the inelastic component, $\eta(t,x)$, in the first-order correction to the solution, $K^{(1)}[w]$ of Eqs. (4.12) and (4.13) is replaced by

$$\begin{aligned} K^{(1)}[w,\eta] = 6a_4 w^2 w_1 - 3ww_3 + \left(\frac{3}{2} - 12a_1 + 6a_4\right) w_1 w_2 + \tfrac{3}{4} w_5 \\ + 6\partial_x(w\eta) + \eta_{xxx} - \eta_t \end{aligned} \quad . \tag{6.3}$$

The first step is to make the $O(\varepsilon)$-part of Eq. (4.12) for $w$ coincide with $O(\varepsilon)$-part of the Normal Form Eq. (2.3). Namely, one wishes to have

$$K^{(1)}[w,\eta] = \alpha_4 S_3[w] \quad . \tag{6.4}$$

$\alpha_4 = (3/4)$ is the coefficient of $w_5$ in Eq. (6.3), (see also Eq. (4.17)), and $S_3$ is a symmetry, given in Eq. (A.2). This requirement determines the dynamical equation obeyed by the inelastic component, $\eta(t,x)$, in lowest order to be:

$$\begin{aligned} \eta_t = 6\partial_x(w\eta) + \eta_{xxx} + U_{\eta,1} + \varepsilon U_{\eta,2} + \varepsilon^2 U_{\eta,3} + O(\varepsilon^3) \\ \left( U_{\eta,1} = \left(6a_4 - \frac{45}{2}\right) w^2 w_1 + \left(6a_4 - 12a_1 - \frac{27}{2}\right) w_1 w_2 - \frac{21}{2} ww_3 \right) \end{aligned} \quad . \tag{6.5}$$

$U_{\eta,2}$ and $U_{\eta,3}$ will be computed in the second-and third-order analyses.

The next step is to make $U_{\eta,1}$ vanish in the single-soliton case. The single-soliton solution is:

$$w(t,x) = 2k^2 / \cosh\left[k(x+vt)\right]^2 \quad . \tag{6.6}$$

Substituting Eq. (6.6) in $U_{\eta,1}$, one finds that this requirement is met for

$$a_1 = -\frac{11}{4} \quad , \quad a_4 = -\frac{3}{2} \quad . \tag{6.7}$$

With this assignment, the first-order elastic term (which is the full first-order contribution in the single-soliton case!) has been determined:

$$w^{(1)} = -\frac{11}{4} w_2 - \frac{3}{2} w^2 \quad , \tag{6.8}$$

and the driving in Eq. (6.5) term becomes

$$U_{\eta,1} = -\frac{21}{2} \partial_x R^{(6,1)} \quad , \quad \left( R^{(6,1)} = w^3 + w w_2 - {w_1}^2 \right) \quad . \tag{6.9}$$

This result has important consequences. First, $R^{(6,1)}$ is a *local special polynomial* of scaling weight 6. (Special polynomials are differential polynomials that vanish identically in the single-soliton case. They are discussed extensively in [20].) Substitution of Eq. (6.6) confirms that $R^{(6,1)}$ vanishes identically in the single-soliton case. When $w$ is a multiple-soliton solution, $R^{(6,1)}$ is localized around the soliton-collision region in the $x - t$ plane and falls off exponentially fast in all directions in the plane away from this region [16, 44, 20]. Furthermore, it does not resonate with the homogeneous part of Eq. (6.5). As a result, the solution for $\eta(t,x)$ vanishes far from soliton trajectories [20]. In fact, with Eq. (6.9) for $U_{\eta,1}$, the lowest-order part of Eq. (6.5) is solved by:

$$\eta(t\,x) = \frac{7}{2} \partial_x \omega(t,x) + H(t,x) \quad , \quad \omega(t,x) = R^{(3,1)} = q w + w_1 \quad . \tag{6.10}$$

In Eq. (6.10), $H(t,x)$ is a solution of the homogeneous part of Eq. (6.5) (A linear combination of symmetries of the KdV equation, see Appendix A.)

The existence of the closed-form solution (Eq. (6.10)) for Eq. (6.5) at the $O(\varepsilon)$ level is nothing more than the known fact, that the PKDVE is integrable through first order [1-7]. However, the structure of $U_{\eta,1}$ (a complete differential of a special polynomial, which is localized around the soliton collision region, and does not resonate with the homogeneous part of Eq. (6.5)) is important. It serves as a guideline for shaping the higher-order driving terms, $U_{\eta,2}$ and $U_{\eta,3}$. The freedom in the expansion allows for their construction so that they have the same properties as $U_{\eta,1}$.

Finally, the differential polynomial $R^{(3,1)}$ in the solution, Eq. (6.10) can be made into a *non-local special polynomial* by a specific choice of the limits of integration in the definition of the non-local quantity $q(t,x)$ (see Eq. (B.1)). As will be shown in the following, this makes $\eta(t,x)$ of Eq. (6.10) into a purely inelastic term. The choice employed is:

$$q(t,x) = \frac{1}{2}\left( \int_{-\infty}^{x} w(t,x)dx - \int_{x}^{+\infty} w(t,x)dx \right) = \int_{-\infty}^{x} w(t,x)dx - \frac{1}{2}\int_{-\infty}^{+\infty} w(t,x)dx \quad . \qquad (6.11)$$

Such a choice is allowed, as it amounts to modifying $\eta(t,x)$ of Eq. (6.10) by the addition of the symmetry, $w_1$, which is a solution of the homogeneous part of Eq. (6.5).

When $w(t,x)$ is the single-soliton solution, $q(t,x)$, as defined in Eq. (6.11), is given by:

$$q(t,x) = 2k\tanh\left[k(x+vt)\right] \quad , \quad q(t,x) \xrightarrow[x\to\pm\infty]{} \pm 2k \quad . \qquad (6.12)$$

When $w$ is an $N$-soliton solution, it asymptotes into a sum of well-separated single-soliton solutions. As a result, the asymptotic form of $q(t,x)$ is:

$$q(t,x) \xrightarrow[\substack{t\to\pm\infty \\ |x|\gg k_N|t|}]{} \pm\sum_{j=1}^{N} 2k_j \quad . \qquad (6.13)$$

With Eq. (6.11), $R^{(3,1)}$ becomes a *non-local special polynomial*: It vanishes identically in the single-soliton case, as direct substitution of Eq. (6.6) shows. However, unlike $R^{(6,1)}$ of Eq. (6.8), $R^{(3,1)}$ is not localized around the soliton-collision region in the multiple-soliton case, owing to the presence of the non-local entity, $q(t,x)$. The asymptotic behavior of $R^{(3,1)}$ is [20]:

$$\omega(t,x) = R^{(3,1)} \xrightarrow[|t|\to\infty]{} \sum_{i=1}^{N} Q_i\, w_{Single}(t,x;k_i) \quad , \quad \left( Q_i = \left\{ \sum_{k_j<k_i} (-2k_j) + \sum_{k_j>k_i} (2k_j) \right\} \mathrm{sgn}(t) \right) \quad .(6.14)$$

(Soliton wave numbers are ordered so that $0 < k_1 < k_2 < \ldots. < k_N$.) In the two-soliton case, Eq. (6.14) becomes

$$\omega(t,x) = R^{(3,1)} \underset{|t|\to\infty}{\to} \left(2k_2\, w_{Single}(t,x;k_1) - 2k_1\, w_{Single}(t,x;k_2)\right)\mathrm{sgn}(t) \quad , \quad (k_2 > k_1) \ . \qquad (6.15)$$

(In Eqs. (6.14) and (6.15), the $w$'s are single-soliton solutions of the Normal Form.) Thus, $R^{(3,1)}$ is a purely inelastic term: Each soliton is multiplied by wave numbers of *other* solitons. In the two-soliton case, $\omega(t,x)$ describes a soliton and an anti soliton, which exchange signs upon an elastic collision (see Fig. 1). Other initial data for $\omega(t,x)$ generate additional inelastic processes [20].

### 6.2 Second-order analysis

Of the 23 monomials allowed in the second-order correction, $w^{(2)}$ (see Eq. B.3), only the four purely local ones are allowed in the elastic component:

$$w^{(2)} = b_1\, w_4 + b_9\, w\, w_2 + b_{10}\, {w_1}^2 + b_{13}\, w^3 \ . \qquad (6.16)$$

For Eq. (4.12) to be a Normal Form through $O(\varepsilon^2)$, one must have (see Eq. (2.3))

$$K^{(2)}[w,\eta] = \beta_8\, S_4[w] \ . \qquad (6.17)$$

In Eq. (6.17), $\beta_8$ = (5/8) is the coefficient of $w_5$ in $K^{(2)}$ (see also Eq. (4.18)), and $S_4$ is a symmetry, given in Eq. (A.2). This requirement determines $U_{\eta,2}$ in Eq. (6.5) to be the remainder of $K^{(2)}$:

$$\begin{aligned} U_{\eta,2} = \partial_x &\left\{ 3\eta^2 + \frac{3}{4}\eta_4 + \frac{9}{2} w_1\, \eta_1 - \frac{39}{2} w_2\, \eta - 3w\eta_2 - 9w^2\, \eta \right\} \\ &+ (7 + 6b_{13})w^3\, w_1 + \left(\frac{275}{4} - 6b_{10} + 6b_{13}\right){w_1}^3 + \left(\frac{1217}{4} - 12b_9 + 18b_{13}\right)w\, w_1\, w_2 + \frac{667}{8} w^2\, w_3 \ . \qquad (6.18) \\ &+ \left(\frac{527}{4} - 60b_1 + 3b_9 + 6b_{10}\right)w_2\, w_3 + \left(\frac{147}{4} - 24b_1 + 3b_9\right)w_1\, w_4 + \frac{125}{8} w\, w_5 \end{aligned}$$

The next step is to make the driving term in $U_{\eta,2}$ vanish for the single-soliton solution of Eq. (6.6). In view of the fact that, in lowest order, Eq. (6.5) is conservative (the lowest-order part of its r.h.s. is a complete differential), it is desirable to exploit the freedom in the expansion to make $U_{\eta,2}$ also into a complete differential. These two requirements are met by assigning the coefficients that appear in Eq. (6.18) the following values:

$$b_1 = \frac{619}{160} \quad , \quad b_9 = \frac{223}{20} \quad , \quad b_{10} = \frac{493}{40} \quad , \quad b_{13} = -\frac{47}{20} \quad . \tag{6.19}$$

With these values of the coefficients, the second-order elastic component, $w^{(2)}$, becomes:

$$w^{(2)} = \frac{619}{160} w_4 + \frac{223}{20} w\, w_2 + \frac{493}{40} w_1^{\,2} - \frac{47}{20} w^3 \quad . \tag{6.20}$$

(This is the full second-order contribution in the single-soliton case!) $U_{\eta,2}$ of Eq. (6.18) becomes

$$\begin{aligned} U_{\eta,2} &= \partial_x \left\{ 3\eta^2 + \frac{3}{4}\eta_4 + \frac{9}{2} w_1 \eta_1 - \frac{39}{2} w_2 \eta - 3 w \eta_2 - 9 w^2 \eta \right\} \\ &+ \partial_x \left\{ -\frac{1879}{20} R^{(8,1)}[w] + \frac{1531}{40} R^{(8,2)}[w] + \frac{125}{8} R^{(8,3)}[w] \right\} \end{aligned} \quad . \tag{6.21}$$

In Eq. (6.21), $R^{(8,k)}$, are the three independent local special polynomials with scaling-weight 8:

$$\begin{aligned} &R^{(8,1)}[w] = w\, R^{(6,1)}[w] \quad , \quad R^{(8,2)}[w] = 2 w^4 - 5 w w_1^{\,2} + 3 w^2 w_2 + w_2^{\,2} - w_1 w_3 \\ &R^{(8,3)}[w] = w^4 + 5 w w_1^{\,2} + 4 w^2 w_2 - w_2^{\,2} + w w_4 \end{aligned} \quad . \tag{6.22}$$

They all share the structure of $R^{(6,1)}$ [20]. Namely, they vanish identically when $w$ is a single-soliton solution. In the multiple-soliton case, they are all localized around the soliton-collision region, and decay exponentially in all directions in the $x - t$ plane.

It is possible to account for the effect of $U_{\eta,2}$ in two ways. One may regard it as a driving term for the next-order contribution to $\eta(t,x)$, or as an additional term in the evolution of $\eta(t,x)$, Eq. (6.5). It is instructive to examine the first option. The contributions of $R^{(8,2)}$ and $R^{(8,3)}$ to the $O(\varepsilon)$-component of $\eta(t,x)$ can be found in closed form:

$$
\begin{aligned}
&\partial_x R^{(8,2)} \to \frac{1}{3}\partial_x\left(-\partial_x{}^2 R^{(3,1)} - 4\,w R^{(3,1)} + q\,\partial_x R^{(3,1)} \ - 3q^2 R^{(3,1)} + w R^{(3,2)}\right) \\
&\partial_x R^{(8,3)} \to \frac{1}{3}\partial_x\left(w R^{(3,1)} - q\,\partial_x R^{(3,1)} \ + 3q^2 R^{(3,1)} - w R^{(3,2)}\right) \qquad . \\
&\left(R^{(3,2)} = 3qw + q^3 - 3q^{(3,1)}\right)
\end{aligned}
\tag{6.23}
$$

Each term in Eq. (6.23) asymptotes into a purely inelastic wave. Consider, for example, the quantity ($w\,R^{(3,2)}$). ($R^{(3,2)}$ is a second *non-local special polynomial* of scaling-weight 3.) When $w$ is a two-soliton solution, ($w\,R^{(3,2)}$) asymptotes into

$$
w R^{(3,2)} \underset{|t|\to\infty}{\to} 6\left\{
\begin{array}{l}
\frac{2}{3}k_2\left(4\,w_{Single}\left(t,x;k_1\right) + \partial_x{}^2 w_{Single}\left(t,x;k_1\right)\right) \\
-\frac{2}{3}k_1\left(4\,w_{Single}\left(t,x;k_2\right) + \partial_x{}^2 w_{Single}\left(t,x;k_2\right)\right) \\
-\,2\,k_1{}^2\,\partial_x w_{Single}\left(t,x;k_2\right) - 2\,k_2{}^2\,\partial_x w_{Single}\left(t,x;k_1\right)
\end{array}
\right\}\mathrm{sgn}(t) \quad , \quad \left(k_2 > k_1\right). \tag{6.24}
$$

On the other hand, $R^{(8,1)}$ is an obstacle to asymptotic integrability [20]. The contribution it generates in the solution cannot be expressed as a differential polynomial in $w$. It has been computed numerically. Denoting this contribution by ($\varepsilon\,\delta\eta$), $\delta\eta$ obeys the equation:

$$
\delta\eta_t = 6\,\partial_x\left(w\,\delta\eta\right) + \delta\eta_{xxx} - \frac{1879}{20}\partial_x R^{(8,1)}\left[w\right] \ . \tag{6.25}
$$

Eq. (6.25) may be simplified by defining

$$
\delta\eta\left(t,x\right) = -\frac{1879}{20}\partial_x\omega_2\left(t,x\right) \ . \tag{6.26}
$$

The equation obeyed by $\omega_2$ is:

$$
\partial_t\omega_2 = 6\,w\partial_x\omega_2 + \partial_x{}^3\omega_2 + R^{(8,1)}\left[w\right] \ . \tag{6.27}
$$

The driving term in Eq. (6.27) is localized around the soliton-collision region, vanishes exponentially away from that region, and does not resonate with the homogeneous part of Eq. (6.27). As a result, $\omega_2$ is expected to be bounded. Hence, like all closed-form contributions to $\eta(t,x)$, $(\varepsilon\ \delta\eta)$, the contribution of the second-order obstacle to integrability, is also expected to be bounded and to vanish away from soliton trajectories.

These expectations are borne out by the numerical solution [20]. The main contribution in $\omega_2$ is similar to the closed-form expressions of Eqs. (6.14), (6.15) and (6.23): A wave that asymptotes into a linear combination of single zero-order solitons solutions of the Normal For Eq. (2.3) [20]. It is followed by a decaying dispersive wave of the type found by [24-26]. An example of the structure of $\omega_2$ is shown in Fig. 2 in the two-soliton case. Zero-initial data have been assumed. Within the numerical accuracy, the leading soliton and anti-soliton have the same velocities, phase shifts and profiles as the zero-order solitons in the two-soliton solution of the Normal Form, Eq. (2.3). Their amplitudes are determined by the driving term in Eq. (6.27). A comparison between the asymptotic form of $\omega_2$ in the two-soliton case and of a sum of two single-solutions (their amplitudes obtained from the numerical solution for $\omega_2$) is shown in Fig. 3.

If one chooses to regard $U_{\eta,2}$ of Eq. (6.21) as another driving term in Eq. (6.5), then thanks to its properties (localized along soliton trajectories; a complete differential with respect to $x$; the driving term it contains is localized around the soliton collision region and falls off exponentially away from that region), for small $\varepsilon$, $U_{\eta,2}$ of Eq. (6.21) will not modify the nature of the solution of Eq. (6.5). The solution will vanish for $|x| \to \infty$ at fixed $t$.

### 6.3 Third-order analysis

Of the 74 monomials allowed in the third-order term, $w^{(3)}$ (see Eq. B.4), only the seven purely local ones are allowed in the elastic component:

$$w^{(3)} = c_1\, w_6 + c_{25}\, w\, w_4 + c_{26}\, w_1\, w_3 + c_{27}\, {w_2}^2 + c_{37}\, w^2\, w_2 + c_{38}\, w\, {w_1}^2 + c_{41}\, w^4 \quad . \qquad (6.28)$$

The first step is to require that Eq. (4.12) is a Normal Form through $O(\varepsilon^3)$, namely, that

$$K^{(3)}\left[w,\eta\right] = \gamma_{14}\, S_5\left[w\right] \quad . \tag{6.29}$$

In Eq. (6.29), $\gamma_{14}$ = (35/64) is the coefficient of $w_7$ in $K^{(3)}$ (see also Eq. (4.19)), and $S_5$ is a symmetry, given in Eq. (A.2). Whatever remains of $K^{(3)}$ is assigned to $U_{\eta,3}$ in Eq. (6.5).

One wishes to make $U_{\eta,3}$ vanish identically in the single-soliton case (so that the driving terms it contains are localized in the $x - t$ plane in the multiple-soliton case), as well as into a complete differential. This determines some of the coefficients $c_k$ in Eq. (6.28). The remaining free coefficients have been exploited so as to simplify $U_{\eta,3}$. The following values have been assigned to $c_k$:

$$\begin{array}{l} c_1 = -\dfrac{11819}{3200}\,,\; c_{25} = -\dfrac{231}{25}\,,\; c_{26} = -\dfrac{3477}{50}\,,\; c_{27} = -\dfrac{24669}{400} \\ c_{37} = \dfrac{13833}{800}\,,\; c_{38} = \dfrac{18873}{400}\,,\; c_{41} = -\dfrac{63}{10} \end{array} \quad , \tag{6.30}$$

The third-order elastic term, $w^{(3)}$, then becomes:

$$\begin{array}{l} w^{(3)} = -\dfrac{11819}{3200} w_6 - \dfrac{231}{25} w\, w_4 - \dfrac{3477}{50} w_1\, w_3 - \dfrac{24669}{400} w_2^{\;2} \\ \quad + \dfrac{13833}{800} w^2\, w_2 + \dfrac{18873}{400} w\, w_1^{\;2} - \dfrac{63}{10} w^4 \end{array} \quad , \tag{6.31}$$

(This is the full third-order term in the single soliton case!). $U_{\eta,3}$ obtains the form

$$U_{\eta,3} = \partial_x \left\{ P\left[w,\eta\right] + Q\left[w\right] \right\} \quad . \tag{6.32}$$

In Eq. (6.32),

$$\begin{array}{l} P\left[w,\eta\right] = \dfrac{75}{4}\eta_1^{\;2} + \dfrac{27}{2}\eta\,\eta_2 + \dfrac{43}{16}\eta_6 + \left( -\dfrac{141}{10} w^3 + \dfrac{267}{10} w_1^{\;2} + \dfrac{177}{5} w\, w_2 - \dfrac{2133}{80} w_4 \right)\eta \\ \quad - \left( 126\, w\, w_1 + \dfrac{855}{8} w_3 \right)\eta_1 - \left( \dfrac{63}{4} w^2 + \dfrac{201}{8} w_2 \right)\eta_2 - \dfrac{15}{4} w_1\, \eta_3 - \dfrac{51}{4} w\, \eta_4 \end{array} \tag{6.33}$$

and

$$Q[w]=\frac{23127}{800}w^5-\frac{547683}{800}w^2w_1^{\,2}-\frac{12297}{80}w^3w_2$$
$$+\frac{849621}{800}w_1^{\,2}w_2-\frac{7617}{100}ww_2^{\,2}+\frac{24639}{320}w^2w_4 \qquad . \qquad (6.34)$$
$$+\frac{216273}{400}ww_1w_2+\frac{19701}{800}w_3^{\,2}-\frac{151317}{1600}w_2w_4+\frac{7461}{160}w_1w_5+\frac{7461}{320}ww_6$$

The driving term, $Q[w]$, is local a special polynomial. It vanishes identically when $w$ is a single-soliton solution. When $w$ is a multiple-soliton solution, $Q[w]$ is localized in the soliton-collision region, and falls off exponentially in all directions in the $x-t$ plane away from that region. Part of $Q[w]$ generates in the inelastic component, $\eta(t,x)$, a closed-form contribution - a differential polynomial in $w$. Part of $Q[w]$ is an obstacle to integrability. The contribution of the latter part to $\eta(t,x)$ has to be found numerically. Both parts generate solutions that asymptote into linear combinations of zero-order solitons, the amplitudes of which are determined by the localized driving terms. The obstacle, in addition, generates a decaying dispersive wave. (See [20] for details.)

**7. Some consequences of physical significance**

To lowest-order, the r.h.s. of Eq. (6.5) is a complete differential:

$$\eta_t=\partial_x\left\{6(w\eta)+\eta_{xx}-\frac{21}{2}\left(w^3+ww_2-w_1^{\,2}\right)\right\}+O(\varepsilon) \quad . \qquad (7.1)$$

The solution of Eq. (7.1) is Eq. (6.10). When $w$ is a multiple-soliton solution of an integrable Normal Form, the particular solution in Eq. (6.10), as well as the symmetries included in $H(t,x)$ all vanish for $|x|\to\infty$ at fixed $t$. As a result, so does the solution of Eq. (7.1). Hence, to lowest order:

$$\frac{d}{dt}\int_{-\infty}^{+\infty}\eta(t,x)dx=0+O(\varepsilon) \quad . \qquad (7.2)$$

Exploiting the freedom, $U_{\eta,2}$ and $U_{\eta,3}$ have been also made into complete differentials, so as to extend the validity of the conservation law, Eq. (7.2), through $O(\varepsilon^2)$:

$$\frac{d}{dt}\int_{-\infty}^{+\infty}\eta(t,x)\,dx = 0 + O(\varepsilon^3) \quad . \tag{7.3}$$

As a result, whether $\eta(t,x)$ is a conserved quantity or not, depends solely on the initial data and/or boundary conditions imposed on Eq. (6.5)!

Next, the elastic component trivially vanishes exponentially fast away from soliton trajectories in the multiple-soliton case. The inelastic component also vanishes there due to the manner of its construction. Hence, the mass conservation law, Eq. (3.4), is obeyed by the perturbative solution. Note that solutions that do not obey the conservation law are possible. However, again, this depends entirely on the initial data and/or boundary conditions imposed on the inelastic component!

Finally, Eq. (3.2) yields the electrostatic potential across the wave solution:

$$\varphi(\tau,\xi) = -\int_{-\infty}^{\xi} v_\tau\,d\xi - \tfrac{1}{2}\left(v^2 - {v_0}^2\right) \quad . \tag{7.4}$$

In terms of re-scaled quantities and coordinates (Eqs. (4.1), (4.5) and (4.10)), and the asymptotic series for the soliton velocity (Eqs. (4.7), (6.1), (6.8), (6.20), and (6.31)), Eq. (7.4) yields:

$$\begin{gathered}\varphi\left(\tau = \frac{t}{2\varepsilon^{3/2}}, \xi = \frac{x}{\varepsilon^{1/2}}\right) = \\ 3\varepsilon\left\{w + \varepsilon\left(3w^2 - \frac{3}{4}w_2\right) + \varepsilon^2\left(-\frac{317}{20}w^3 + \frac{433}{40}{w_1}^2 - \frac{128}{5}w\,w_2 - \frac{21}{160}w_4\right)\right\} \quad . \\ 3\varepsilon^2\left\{\eta + \varepsilon\left(9w\eta + 2\eta_2\right)\right\} + O(\varepsilon^4)\end{gathered} \tag{7.5}$$

Here $w = w(t,x)$ and $\eta = \eta(t,x)$. Thus, the electrostatic potential is a first-order quantity. The inelastic component affects it only in the second order. As a result, to lowest-order, the electrostatic potential is linear in $w$, the solution of the Normal Form, Eq. (2.3), with $\alpha_4 = (3/4)$, $\beta_8 = (5/8)$ and $\gamma_{14} = (35/64)$. Namely, it is approximately proportional to the ion velocity wave.

Up to the numerical factor, $3\varepsilon$, in the scaled coordinates, the charge density is given in lowest order by

$$\rho = -w_{xx} \quad . \tag{7.6}$$

Fig.4 shows this approximation for $\rho$ across a single-soliton solution. It describes a symmetric triple layer. In the case of a multiple-soliton solution, Fig. 4 describes the density profile across each soliton sufficiently far from the soliton-collision region. Fig. 5 shows the charge density across the soliton-collision region for a two-soliton solution. It is a triple layer, but not symmetric.

**8. Summary**

In this paper, the perturbative analysis of the ion acoustic wave equations of Plasma Physics has been addressed. The freedom in the expansion allows for the avoidance of pure non-local terms that do not vanish away from soliton trajectories (they are unavoidable in the shallow-water problem), but does not help to eliminate obstacles to asymptotic integrability. As obstacles spoil the simplicity of the zero-order term in the traditional analysis, a new expansion scheme has been employed, in which the elastic component is constructed so as to have the same structure as the full solution in the single-soliton case (where obstacles to integrability do not emerge, and non-local terms can be totally avoided). The zero-order approximation of the elastic component is the solution of an asymptotically integrable Normal Form.

The inelastic component represents the net effect of soliton interactions in the multiple-soliton case. Eq. (6.5) governs the evolution of the inelastic term. It is driven by terms that are localized in the soliton-collision region. All the driving terms on the r.h.s. of Eq. (6.5) (integrable ones and obstacles) are all localized around the origin in the $x - t$ plane, fall off exponentially away from there, and do not resonate with the homogeneous part of Eq. (6.5). This results in a solution for $\eta(t,x)$ that is bounded and vanishes away from soliton trajectories. Through third order, the leading part of this solution is a wave of solitons and anti-solitons that are asymptotically proportional to the zero-order single-soliton solutions of the Normal Form [20]. Hence, they do not lead to modification of soliton parameters. The obstacles add a decaying dispersive wave. The emerging physical picture is in agreement with the numerical results of [19].

The inelastic component is constructed in such a manner that physically meaningful properties of the full solution depend solely on the boundary conditions and/or initial data chosen for the inelastic component. This includes the satisfaction by the inelastic component of the conservation law, Eq. (7.3), and of mass conservation Eq. (3.4) by the disturbance to the ion density.

Finally, the two-component approach can be applied to other dynamical systems, the solutions of which can be described approximately by solitons solutions of integrable non-linear wave equations. This is the case with the classical shallow-water problem, as well as with systems that can be approximated by the NLS equation. The results of these analyses will be published elsewhere.

Acknowledgment: Invaluable discussions with G. I. Burde are acknowledged.

**Appendix A. Symmetries**

The equation that determines $u^{(n)}$, the $n^{\text{th}}$-order correction in the solution of the PKDVE is

$$\partial_t u^{(n)} = 6\partial_x\left(u\,u^{(n)}\right) + \partial_x^{\,3} u^{(n)} + F^{(n)}[u] \quad . \tag{A.1}$$

$F^{(n)}$ represents the effect of the $n^{\text{th}}$-order perturbation in Eq. (2.1) and the lower-order corrections, $u^{(k)}$, $1 \le k \le n$-1. The homogeneous part of Eq. (A.1), the linearization of the KdV equation, is

$$\partial_t S[u] = 6\partial_x\left(u\,S\right) + \partial_x^{\,3} S \quad . \tag{A.2}$$

Eq. (A.2) has an infinite hierarchy of solutions [3-7, 37-56], the symmetries of the KdV equation. The first few are:

$$\begin{gathered} S_1 = u_x \\ S_2 = 6u\,u_1 + u_3 \\ S_3 = 30u^2 u_1 + 10u\,u_3 + 20u_1 u_2 + u_5 \\ S_4 = 140u^3 u_1 + 70u^2 u_3 + 280u\,u_1 u_2 + 14u\,u_5 + 70u_1^{\,3} + 42u_1 u_4 + 70u_2 u_3 + u_7 \\ S_5 = 630u^4 u_1 + 1260u\left(u_1\right)^3 + 2520u^2 u_1 u_2 + 1302u_1\left(u_2\right)^2 + 420u^3 u_3 + 966\left(u_1\right)^2 u_3 \\ \hfill \left(u_p \equiv \partial_x^{\,p} u\right) \end{gathered} \quad . \tag{A.3}$$

As a symmetry is a solution of Eq. (A.2), adding a symmetry as a driving term:

$$\partial_t u^{(n)} = 6\partial_x\left(u\,u^{(n)}\right) + \partial_x^{\,3} u^{(n)} + \mu\,S_n[u] \quad , \tag{A.5}$$

will generate a linearly unbounded solution for $u^{(n)}(t,x)$. This is resolved by shifting the contribution of symmetries to the Normal Form, Eq. (2.3).

**Appendix B. Differential polynomials for higher-order corrections**

The differential polynomials given below are to be used in the perturbative analysis of the solution of Eq. (2.1). Differential polynomials of identical composition are to be used in Sections 3 - 6, where the derivation of Eq. (2.1) for the ion acoustic wave equations is presented.

A convenient manner to generate all differential monomials allowed in a given order is the use of the notion of scaling weights [37-43]. If $u$, $\partial_t$ and $\partial_x$ are assigned the weights 2, 3 and 1, respec-tively, then all terms in the KdV equation have a cumulative scaling weight of 5. The higher-order perturbation terms have scaling weight of (2 $n$ + 5), where $n$ is the order of the perturbation term. The analysis then dictates that $u^{(n)}$, the $n^{\text{th}}$-order correction to the solution ought to have scaling weight of 2($n$ + 1). The monomials may be constructed from powers of $u$ and its derivatives with respect to $x$, as well non-local terms that are integrals of such monomials with respect to $x$. Considering the application to soliton solutions, only non-local terms that are bounded in the $x$ - $t$ plane for such solutions are incorporated. For an analysis through third order, these are:

$$\begin{array}{l}
q=\partial_x^{-1}w \quad , \quad q^{(3,1)}(t,x)=\partial_x^{-1}\left(w^2\right) \quad , \quad q^{(4,1)}(t,x)=\partial_x^{-1}\left(w^2 q\right)\\
q^{(5,1)}(t,x)=\partial_x^{-1}\left(w^3\right) \quad , \quad q^{(5,2)}(t,x)=\partial_x^{-1}\left(w^2 q^2\right) \quad , \quad q^{(5,3)}(t,x)=\partial_x^{-1}\left(w_1^{\,2}\right)\\
q^{(6,1)}(t,x)=\partial_x^{-1}\left(w^3 q\right) \quad , \quad ,q^{(6,2)}(t,x)=\partial_x^{-1}\left(w^2 q^3\right) \quad , \quad q^{(6,3)}(t,x)=\partial_x^{-1}\left(w_1^{\,2} q\right)\\
q^{(7,1)}(t,x)=\partial_x^{-1}\left(w^4\right) \quad , \quad q^{(7,2)}(t,x)=\partial_x^{-1}\left(w^2 q^2\right) \quad , \quad q^{(7,3)}(t,x)=\partial_x^{-1}\left(w^2 q^4\right)\\
q^{(7,4)}(t,x)=\partial_x^{-1}\left(w w_1^{\,2}\right) \quad , \quad q^{(7,5)}(t,x)=\partial_x^{-1}\left(w_1^{\,2} q^2\right) \quad , \quad q^{(7,6)}(t,x)=\partial_x^{-1}\left(w_2^{\,2}\right)\\
q^{(7,7)}(t,x)=\partial_x^{-1}\left(w^2 q^{(4,1)}\right)\\
q^{(8,1)}(t,x)=\partial_x^{-1}\left(w^4 q\right) \quad , \quad q^{(8,2)}(t,x)=\partial_x^{-1}\left(w^3 q^3\right) \quad , \quad q^{(8,3)}(t,x)=\partial_x^{-1}\left(w^2 q^5\right)\\
q^{(8,4)}(t,x)=\partial_x^{-1}\left(w w_1^{\,2} q\right) \quad , \quad q^{(8,5)}(t,x)=\partial_x^{-1}\left(w_1^{\,2} q^3\right) \quad , \quad q^{(8,6)}(t,x)=\partial_x^{-1}\left(w_2^{\,2} q\right)\\
q^{(8,7)}(t,x)=\partial_x^{-1}\left(w_1^{\,3}\right) \quad , \quad q^{(8,8)}(t,x)=\partial_x^{-1}\left(w^3 q^{(3,1)}\right) \quad , \quad q^{(8,9)}(t,x)=\partial_x^{-1}\left(w^2 q^{(5,2)}\right)\\
q^{(8,10)}(t,x)=\partial_x^{-1}\left(w^2 q^{(5,3)}\right) \quad , \quad q^{(8,11)}(t,x)=\partial_x^{-1}\left(w_1^{\,2} q^{(3.1)}\right)
\end{array} \quad . \tag{B.1}$$

In $q^{(k,l)}$, $k$ denotes the scaling weight, and $l$ is a running index. Scaling-weight considerations allow the following expressions for $w^{(n)}$, with $w_k = \partial_x^{\,k} w$:

$$w^{(1)} = a_1 w_2 + a_2 w_1 q + a_3 w q^2 + a_4 w^2 + a_5 q^4 + a_6 q^{(3,1)} q + a_7 q^{(4,1)} \quad , \tag{B.2}$$

$$\begin{array}{l}
w^{(2)} = b_1 w_4 + b_2 w_3 q + b_3 w_2 q^2 + b_4 w_1 q^3 + b_5 w_1 q^{(3)} + b_6 w q^4 + b_7 w q q^{(3,1)} + b_8 w q^{(4,1)}\\
\qquad + b_9 w w_2 + b_{10} w_1^{\,2} + b_{11} w w_1 q + b_{12} w^2 q^2 + b_{13} w^3\\
\qquad + b_{14} q^{(6,1)} + b_{15} q^{(6,2)} + b_{16} q^{(6,3)} + b_{17} q q^{(5,1)} + b_{18} q q^{(5,2)} + b_{19} q q^{(5,3)} + b_{20} q^2 q^{(4,1)}\\
\qquad + b_{21} q^3 q^{(3,1)} + b_{22} \left(q^{(3,1)}\right)^2 + b_{23} q^6
\end{array} \quad , \tag{B.3}$$

$$
\begin{aligned}
w^{(3)} = {} & c_1 w_6 + c_2 w_5 q + c_3 w_4 q^2 + c_4 w_3 q^3 + c_5 w_3 q^{(3)} + c_6 w_2 q^4 + c_7 w_2 q q^{(3)} + c_8 w_2 q^{(4)} \\
& + c_9 w_1 q^5 + c_{10} w_1 q^2 q^{(3,1)} + c_{11} w_1 q q^{(4,1)} + c_{12} w_1 q^{(5,1)} + c_{13} w_1 q^{(5,2)} + c_{14} w_1 q^{(5,3)} \\
& + c_{15} w q^6 + c_{16} w q^3 q^{(3,1)} + c_{17} w q^2 q^{(4,1)} + c_{18} w q q^{(5,1)} + c_{19} w q q^{(5,2)} + c_{20} w q q^{(5,3)} + c_{21} w q^{(6,1)} \\
& + c_{22} w q^{(6,2)} + c_{23} w q^{(6,3)} + c_{24} w \left(q^{(3,1)}\right)^2 + c_{25} w w_4 + c_{26} w_1 w_3 + c_{27} \left(w_2\right)^2 + c_{28} w w_3 q \\
& + c_{29} w_1 w_2 q + c_{30} w w_2 q^2 + c_{31} w_1^2 q^2 + c_{32} w w_1 q^3 + c_{33} w w_1 q^{(3,1)} + c_{34} w^2 q^4 \\
& + c_{35} w^2 q q^{(3,1)} + c_{36} w^2 q^{(4,1)} + c_{37} w^2 w_2 + c_{38} w w_1^2 + c_{39} w^2 w_1 q + c_{40} w^3 q^2 + c_{41} w^4 \\
& + c_{42} q^{(8,1)} + c_{43} q^{(8,2)} + c_{44} q^{(8,3)} + c_{45} q^{(8,4)} + c_{46} q^{(8,5)} + c_{47} q^{(8,6)} + c_{48} q^{(8,7)} + c_{49} q^{(8,8)} \\
& + c_{50} q^{(8,9)} + c_{51} q^{(8,10)} + c_{52} q^{(8,11)} + c_{53} q q^{(7,1)} + c_{54} q q^{(7,2)} + c_{55} q q^{(7,3)} + c_{56} q q^{(7,4)} \\
& + c_{57} q q^{(7,5)} + c_{58} q q^{(7,6)} + c_{59} q q^{(7,7)} + c_{60} q q^{(3,1)} q^{(4,1)} + c_{61} q^2 q^{(6,1)} + c_{62} q^2 q^{(6,2)} \\
& + c_{63} q^2 q^{(6,3)} + c_{64} q^2 \left(q^{(3,1)}\right)^2 + c_{65} q^3 q^{(5,1)} + c_{66} q^3 q^{(5,2)} + c_{67} q^3 q^{(5,3)} + c_{68} q^4 q^{(4,1)} \\
& + c_{69} q^5 q^{(3,1)} + c_{70} q^{(3,1)} q^{(5,1)} + c_{71} q^{(3,1)} q^{(5,2)} + c_{72} q^{(3,1)} q^{(5,3)} + c_{73} \left(q^{(4,1)}\right)^2 + c_{74} q^8
\end{aligned}
\qquad .\text{(B.4)}
$$

**Appendix C. Conditions for absence of third-order obstacles to integrability**

Extending the analysis of [3-7] to third order, it is found that for the absence of third-order obstacles to asymptotic integrability, the following quantities must vanish.

$$
\begin{aligned}
\mu_{31} = {} & \tfrac{100}{3}\alpha_1^2\alpha_2 + \tfrac{325}{9}\alpha_1\alpha_2^2 + \tfrac{500}{27}\alpha_2^3 - 200\alpha_1^2\alpha_3 + \tfrac{4300}{9}\alpha_1\alpha_2\alpha_3 + \tfrac{5500}{27}\alpha_2^2\alpha_3 - \tfrac{500}{3}\alpha_1\alpha_3^2 \\
& - \tfrac{2000}{9}\alpha_2\alpha_3^2 + \tfrac{2000}{27}\alpha_3^3 + 150\alpha_1^2\alpha_4 - \tfrac{2930}{9}\alpha_1\alpha_2\alpha_4 - \tfrac{12895}{27}\alpha_2^2\alpha_4 + \tfrac{1030}{9}\alpha_1\alpha_3\alpha_4 \\
& - \tfrac{2600}{9}\alpha_2\alpha_3\alpha_4 + \tfrac{1040}{9}\alpha_3^2\alpha_4 - \tfrac{260}{3}\alpha_1\alpha_4^2 + \tfrac{20210}{27}\alpha_2\alpha_4^2 - \tfrac{1840}{9}\alpha_4^3 \\
& + \left(140\alpha_1 - 35\alpha_2 + \tfrac{140}{3}\alpha_3 - \tfrac{532}{3}\alpha_4\right)\beta_1 + \left(-\tfrac{455}{3}\alpha_1 - \tfrac{175}{3}\alpha_2 - \tfrac{1610}{9}\alpha_3 - \tfrac{3773}{9}\alpha_4\right)\beta_2 \\
& + \left(-560\alpha_1 - 280\alpha_2 + \tfrac{980}{3}\alpha_3 + \tfrac{5026}{9}\alpha_4\right)\beta_3 + \left(\tfrac{980}{3}\alpha_1 + \tfrac{4655}{9}\alpha_2 - \tfrac{1960}{9}\alpha_3 - \tfrac{6314}{9}\alpha_4\right)\beta_4 \\
& + \left(420\alpha_1 + \tfrac{175}{3}\alpha_2 - 140\alpha_3 - \tfrac{1078}{3}\alpha_4\right)\beta_5 + \left(-455\alpha_1 - \tfrac{770}{3}\alpha_2 + 420\alpha_3 + 364\alpha_4\right)\beta_6 \\
& + \left(315\alpha_1 + \tfrac{350}{3}\alpha_2 - \tfrac{700}{3}\alpha_3 - \tfrac{728}{3}\alpha_4\right)\beta_7 + \left(-\tfrac{140}{3}\alpha_1 - \tfrac{700}{9}\alpha_2 + \tfrac{1288}{9}\alpha_4\right)\beta_8 \\
& - 63\gamma_1 - 189\gamma_2 + 294\gamma_3 + \tfrac{1302}{5}\gamma_4 + 105\gamma_5 - \tfrac{1932}{5}\gamma_6 - 189\gamma_7 + 441\gamma_8 \\
& + 252\gamma_9 - 357\gamma_{10} - 504\gamma_{11} + 336\gamma_{12}
\end{aligned}
\qquad , \quad \text{(C.1)}
$$

$$
\begin{aligned}
\mu_{32} = &-\tfrac{350}{3}\alpha_1^2\alpha_2 - \tfrac{1100}{9}\alpha_1\alpha_2^2 - \tfrac{2050}{27}\alpha_2^3 + 700\alpha_1^2\alpha_3 - \tfrac{14300}{9}\alpha_1\alpha_2\alpha_3 - \tfrac{24500}{27}\alpha_2^2\alpha_3 + 600\alpha_1\alpha_3^2 \\
&+ \tfrac{5600}{9}\alpha_2\alpha_3^2 - \tfrac{4000}{27}\alpha_3^3 - 500\alpha_1^2\alpha_4 + \tfrac{31015}{36}\alpha_1\alpha_2\alpha_4 + \tfrac{115555}{54}\alpha_2^2\alpha_4 - \tfrac{9295}{18}\alpha_1\alpha_3\alpha_4 \\
&+ \tfrac{16325}{9}\alpha_2\alpha_3\alpha_4 - \tfrac{6530}{9}\alpha_3^2\alpha_4 + \tfrac{3265}{6}\alpha_1\alpha_4^2 - \tfrac{386755}{108}\alpha_2\alpha_4^2 + \tfrac{9005}{9}\alpha_4^3 \\
&+ \left(-490\alpha_1 + 140\alpha_2 - \tfrac{700}{3}\alpha_3 + \tfrac{8351}{12}\alpha_4\right)\beta_1 + \left(\tfrac{1540}{3}\alpha_1 + \tfrac{1435}{6}\alpha_2 + \tfrac{7945}{9}\alpha_3 - \tfrac{31577}{18}\alpha_4\right)\beta_2 \\
&+ \left(1820\alpha_1 + 1225\alpha_2 - \tfrac{3080}{3}\alpha_3 - \tfrac{41839}{18}\alpha_4\right)\beta_3 + \left(-\tfrac{5075}{6}\alpha_1 - \tfrac{42455}{18}\alpha_2 + \tfrac{980}{9}\alpha_3 + \tfrac{124747}{36}\alpha_4\right)\beta_4 \\
&+ \left(-1400\alpha_1 - \tfrac{1435}{6}\alpha_2 + 280\alpha_3 + \tfrac{4571}{3}\alpha_4\right)\beta_5 + \left(1330\alpha_1 + \tfrac{3430}{3}\alpha_2 - 560\alpha_3 - \tfrac{4571}{2}\alpha_4\right)\beta_6 \\
&+ \left(-910\alpha_1 - 840\alpha_2 + \tfrac{1400}{3}\alpha_3 + \tfrac{4571}{3}\alpha_4\right)\beta_7 + \left(\tfrac{70}{3}\alpha_1 + \tfrac{13405}{18}\alpha_2 - \tfrac{7721}{9}\alpha_4\right)\beta_8 \\
&+ 252\gamma_1 + 882\gamma_2 - 1344\gamma_3 - \tfrac{2604}{5}\gamma_4 - \tfrac{861}{2}\gamma_5 + \tfrac{5313}{10}\gamma_6 + 1197\gamma_7 - 1638\gamma_8 \\
&- 504\gamma_9 + \tfrac{3129}{2}\gamma_{10} + 672\gamma_{11} - 168\gamma_{12} - \tfrac{1197}{2}\gamma_{13} + 105\gamma_{14}
\end{aligned}
\quad ,\text{(C.2)}
$$

$$
\begin{aligned}
\mu_{33} = &-\tfrac{40}{3}\alpha_1^2\alpha_2 - \tfrac{100}{9}\alpha_1\alpha_2^2 - \tfrac{290}{27}\alpha_2^3 + 80\alpha_1^2\alpha_3 - \tfrac{1120}{9}\alpha_1\alpha_2\alpha_3 - \tfrac{5500}{27}\alpha_2^2\alpha_3 + 80\alpha_1\alpha_3^2 \\
&- \tfrac{320}{9}\alpha_2\alpha_3^2 + \tfrac{1600}{27}\alpha_3^3 - 40\alpha_1^2\alpha_4 - \tfrac{1621}{36}\alpha_1\alpha_2\alpha_4 + \tfrac{22463}{54}\alpha_2^2\alpha_4 - \tfrac{2627}{18}\alpha_1\alpha_3\alpha_4 \\
&+ \tfrac{5545}{9}\alpha_2\alpha_3\alpha_4 - \tfrac{2218}{9}\alpha_3^2\alpha_4 + \tfrac{1109}{6}\alpha_1\alpha_4^2 - \tfrac{81503}{108}\alpha_2\alpha_4^2 + \tfrac{1753}{9}\alpha_4^3 \\
&+ \left(-56\alpha_1 + 28\alpha_2 - \tfrac{224}{3}\alpha_3 + \tfrac{8183}{60}\alpha_4\right)\beta_1 + \left(\tfrac{140}{3}\alpha_1 + \tfrac{203}{6}\alpha_2 + \tfrac{2177}{9}\alpha_3 - \tfrac{29561}{90}\alpha_4\right)\beta_2 \\
&+ \left(112\alpha_1 + 273\alpha_2 - \tfrac{112}{3}\alpha_3 - \tfrac{49567}{90}\alpha_4\right)\beta_3 + \left(\tfrac{413}{6}\alpha_1 - \tfrac{8659}{18}\alpha_2 - \tfrac{2156}{9}\alpha_3 + \tfrac{142051}{180}\alpha_4\right)\beta_4 \\
&+ \left(-112\alpha_1 - \tfrac{203}{6}\alpha_2 - 112\alpha_3 + \tfrac{5662}{15}\alpha_4\right)\beta_5 + \left(-28\alpha_1 + \tfrac{770}{3}\alpha_2 + 392\alpha_3 - \tfrac{7763}{10}\alpha_4\right)\beta_6 \\
&+ \left(28\alpha_1 - 252\alpha_2 - \tfrac{560}{3}\alpha_3 + \tfrac{7763}{15}\alpha_4\right)\beta_7 + \left(-\tfrac{151}{3}\alpha_1 + \tfrac{3353}{18}\alpha_2 - \tfrac{7553}{45}\alpha_4\right)\beta_8 \\
&+ \tfrac{252}{5}\gamma_1 + 252\gamma_2 - \tfrac{1848}{5}\gamma_3 + \tfrac{5208}{25}\gamma_4 - \tfrac{609}{10}\gamma_5 - \tfrac{17871}{50}\gamma_6 + 315\gamma_7 - 252\gamma_8 \\
&+ \tfrac{1008}{5}\gamma_9 + \tfrac{3129}{10}\gamma_{10} - \tfrac{2352}{5}\gamma_{11} + \tfrac{1848}{5}\gamma_{12} - \tfrac{441}{2}\gamma_{13} + 21\gamma_{14}
\end{aligned}
\quad .\text{(C.3)}
$$

57. Figure Captions

Fig. 1 Asymptotic behavior of $\omega$ of Eq. (6.10) in two-soliton case, long before ($t = -800$) and after ($t = +800$) collision, $k_1 = 0.1$, $k_2 = 0.2$.

Fig. 2 Solution of Eq. (6.27) for $\omega_2$, in two-soliton case, $k_1 = 0.1$, $k_2 = 0.2$.

Fig.3 Comparison of asymptotic behavior of $\omega_2$ of Eq. (6.27) with single-soliton profiles in two-soliton case, $k_1 = 0.1$, $k_2 = 0.2$. Full line - $\omega_2$, dashed line – sum of two single solitons with amplitudes determined from numerical solution for $\omega_2$.

Fig. 4 Lowest-order approximation to charge density across single-soliton solution, $k = 0.2$.

Fig. 5 Lowest-order approximation to charge density across two-soliton solution within soliton-collision region ($t = 0$), $k_1 = 0.1$, $k_2 = 0.2$.

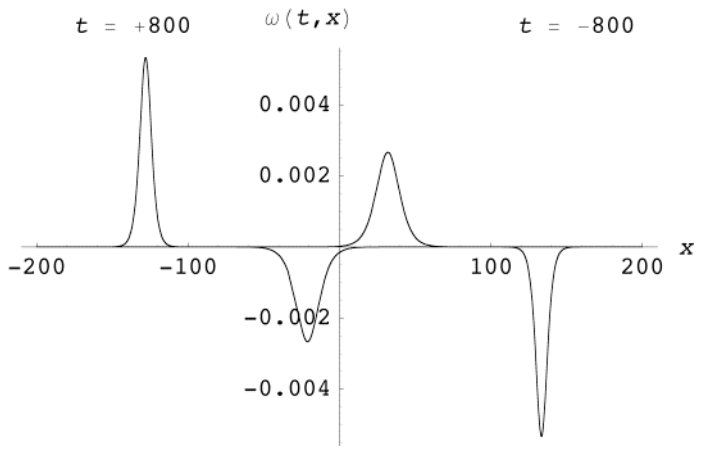

Fig. 1

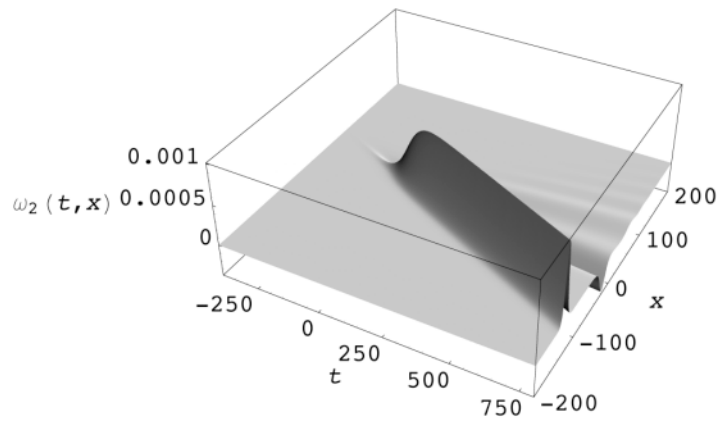

Fig. 2

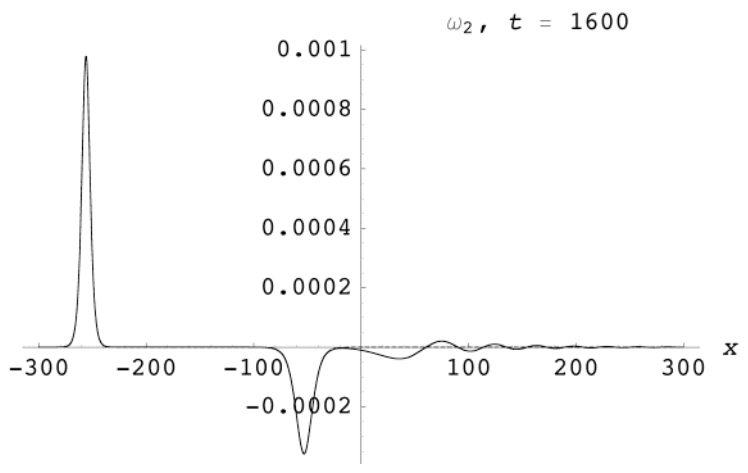

Fig. 3

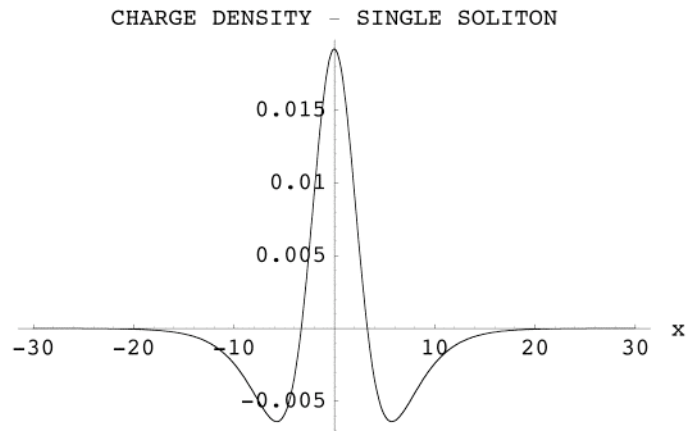

Fig. 4

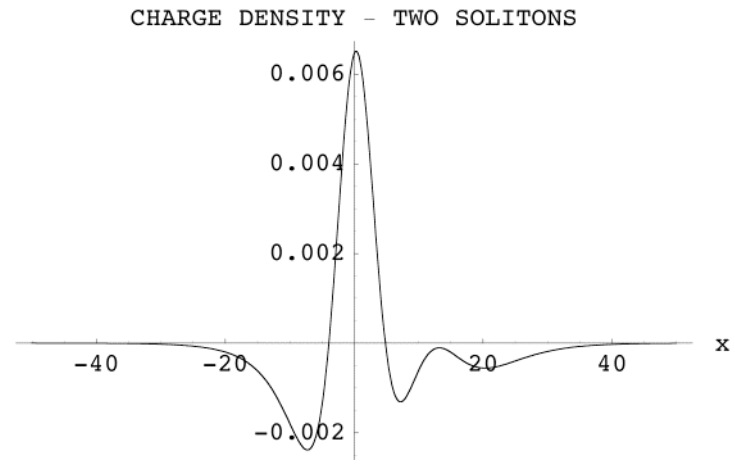

Fig. 5